\documentclass[pre,twocolumn,a4paper]{revtex4}
\usepackage[dvips]{graphicx}
\usepackage{amsfonts}
\usepackage{amsmath}
\usepackage{amssymb}
\usepackage[latin1]{inputenc}

\newcommand{\nnn}{\textbf{n}}
\newcommand{\nablabf}{\boldsymbol{\nabla}}
\newcommand{\pn}{p^{{}}_n}
\newcommand{\on}{\omega^{{}}_n}
\newcommand{\onsqr}{\omega^{2_{}}_n}

\begin{document}

\title{Acoustic resonances in microfluidic chips: \\full-image
 micro-PIV experiments and numerical simulations}

\author{S. M. Sundin, T. Glasdam Jensen, H. Bruus and J. P. Kutter}

\affiliation{MIC -- Department of Micro and Nanotechnology,
Technical University of Denmark\\
DTU Bldg.\ 345 east, DK-2800 Kongens Lyngby, Denmark}

\date{30 March 2007}

\begin{abstract}
We show that full-image micro-PIV analysis in combination with
images of transient particle motion is a powerful tool for
experimental studies of acoustic radiation forces and acoustic
streaming in microfluidic chambers under piezo-actuation in the MHz
range. The measured steady-state motion of both large 5~$\mu$m and
small 1~$\mu$m particles can be understood in terms of the acoustic
eigenmodes or standing ultra-sound waves in the given experimental
microsystems. This interpretation is supported by numerical
solutions of the corresponding acoustic wave equation.
\end{abstract}

\maketitle

\section{Introduction}
\label{sec:Intro}

For the typical dimensions of microfluidic structures there are two
acoustic effects of main importance: \emph{the acoustic radiation
force}~\cite{King1934,Yosioka1955,Gorkov1962}, which moves suspended
particles either towards or away from pressure nodes depending on
their acoustic material properties, and \emph{acoustic
streaming}~\cite{Rayleigh1883,Riley2001}, which imparts movement
onto the entire solvent. Both of these forces have been utilized,
alone or in combination, for several lab-on-a-chip applications.
Yasuda \emph{et al.}~\cite{Yasuda1995,Yasuda1996}, demonstrated
concentration of particles by acoustic radiation forces, and
separation of particles by acoustic forces in combination with
electrostatic forces. One of the most attractive applications for
acoustics in microfluidics is for
mixing~\cite{Zhu1998,Yang2001,Liu2003a}, as this process typically
is diffusion limited in microscale devices. Valveless ultrasonic
pumps, utilizing acoustic streaming, have also been
presented~\cite{Rife2000,Andersson2001}. Numerous examples of
microsystems where acoustics are applied to handling and analysis of
biological material have been suggested. Among others these include:
trapping of microorganisms~\cite{Saito2002},
bioassays~\cite{Lilliehorn2005}, and separation and cleaning of
blood~\cite{Nilsson2004,Jonsson2004,Li2004}. Apart from on chip
devices, acoustic forces have also been suggested for use in other
$µ$m-scale applications~\cite{Wiklund2003}.

There are different imaging strategies and tools, which can be used
in order to enhance the understanding, and to visualize the function
of acoustic micro-devices during operation. For acoustic mixers the
effect can be illustrated and measured by partly filling the mixing
chamber with a dye prior to
piezo-actuation~\cite{Yang2001,Liu2003a}. However, this approach is
mainly limited to determine the total, and not the local, mixing
behavior within the chamber. A more refined method, which is not
limited to the study of micromixers, is to apply streak- or
streamline analysis. This was shown by Lutz \emph{et
al.}~\cite{Lutz2003,Lutz2005}, who neatly demonstrated 3D steady
micro streaming around a cylinder. Although streamline analysis can
be employed to illustrate flow behaviour, it is not suitable in
determining local variations in velocity. For that purpose, the
micron-resolution particle image velocimetry (micro-PIV) technique
is the method of choice~\cite{Santiago1998}. With this technique the
motion of tracer particles, acquired from consecutive image frames,
is utilized to obtain velocity vector fields. In a large chamber,
local measurements of particle motion induced by acoustic radiation
forces and acoustic streaming have been performed by Spengler
\emph{et al.}~\cite{Spengler2000a,Spengler2003}, and further
developed by Kuznetsova \emph{et al.}~\cite{Kuznetsova2004}. Li and
Kenny derived velocity profiles in a particle separating device
utilizing the acoustic radiation force~\cite{Li2004}. Jang \emph{et
al.} used confocal scanning microscopy to perform micro-PIV
measurements on circulatory flows in a piezo-actuated fluidic
chamber~\cite{Jang2005}. Furthermore, Manasseh \emph{et al.} applied
micro-PIV to measure streaming velocites around a bubble trapped in
a microfluidic chamber~\cite{Manasseh2006}.

As particles under the influence of acoustic fields do no longer
function as true independent tracers in all situations, and as
several acoustic effects come into play at the same time, extra
caution and consideration have to be taken when applying micro-PIV
for microfluidic acoustic studies. These considerations will be
discussed in more detail in section \ref{sec:MicroPIV}. The
situation is further complicated by the coupling from the actuator
to the structures and their acoustic resonances, which is a yet
poorly understood mechanism. The resonances depend on the acoustic
material parameters as well as the geometry of both the chip and the
chamber. For substrate materials with low attenuation, such as
silicon, the actuation will result in strong resonances over the
whole devices, whereas for substrate materials with high
attenuation, the effect will be mostly confined to the proximity of
the actuator. Moreover, in a real system the coupling strengths vary
for different resonances, and amplitude fluctuations across the
structures are often observed. Therefore, if investigations striving
to yield a better understanding of acoustic resonances in low
attenuation microfluidic chips are to be performed, it is not
sufficient only to study the acoustic phenomena locally.

In this work, full-image micro-PIV analysis in combination with
images of transient particle motion is suggested as a tool for
studying acoustic resonances in microfluidic chambers under
piezo-actuation. The acousto-fluidic phenomena mentioned above can
be investigated by comparing these experimental images with plots of
acoustic eigenmodes of the device structure obtained by numerical
solution of the corresponding acoustic wave equation.

\section{Materials and experimental methods}
\label{sec:MaterialExp}

\subsection{Microchip fabrication}
\label{sec:ChipFab}

\begin{figure}
\includegraphics[width=0.92\columnwidth]{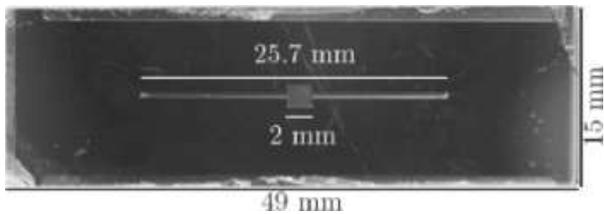}
\caption{A top-view photograph of the silicon-glass chip (dark gray)
containing a square chamber with straight inlet and outlet channels
(light gray).} \label{fig:ChipPhoto}
\end{figure}

In this study, two microfluidic chambers were investigated, one of
quadratic footprint with a side-length of 2~mm and one of circular
shape with a diameter of 2~mm. The size was chosen to be a few times
the acoustic wavelength of 2~MHz ultrasound waves in water, and the
specific shapes were employed to ensure simple patterns in the
pressure field at the acoustic resonances. Both chambers were
connected to 400~$\mu$m wide inlet and outlet channels, and the
depth was 200~$\mu$m throughout. The microfluidic chips were
fabricated in silicon via deep reactive ion etching (DRIE). The same
technique was also applied on the backside of the chip to etch
300~$\mu$m diameter round fluidic inlets. Anodic bonding was used to
seal the structures with a 500~$\mu$m thick pyrex glass lid on the
channel side. Silicon rubber tubings were glued to the holes on the
backside of the chip, for easy attachment of teflon tubing. A
picture of one of our microfluidic chips is shown in
Fig.~\ref{fig:ChipPhoto}, and a list of the geometrical parameters
is given in Table~\ref{tab:chipgeom}.

\subsection{Experimental setup and procedure}
\label{sec:ExpSetup}

The piezo-actuator (Pz27, Ferroperm) was pressed to the backside of
the chip using an ultrasonic gel (ECO, Ceracarta) and biased by a
20~V ac tone generator (Model 195, Wavetek). Images were captured
with a progressive scan interline CCD camera (Hisense MkII, Dantec
Dynamics), mounted with a 0.63x TV-adapter on an epifluorescent
microscope (DMLB, Leica). The objective used was a Plan 5x with a
numerical aperture NA of 0.12. For the given fluidic geometries,
this combination allowed capture of full-image PIV vector fields,
while utilizing the largest number of pixels on the CCD. A blue
light emitting diode, LED, (Luxeon Star 3W, Lumileds) was used as
illumination source in a front-lit configuration, which is described
elsewhere~\cite{Sundin2007}. The LED was powered by an in-house
built power supply controlled by a PIV timing system (Dantec
Dynamics). Image acquisition was performed on a PC with Flowmanager
software (Dantec Dynamics). As tracer fluids solutions of 1~$\mu$m
polystyrene micro-beads (Duke Scientific), 5~$\mu$m polyamide
micro-beads (Danish Phantom Design), diluted milk, and fluorescein
have been used.

The investigations were performed by scanning the applied
frequency from the tone generator and identifying those
frequencies which led to a strong response, an acoustic resonance,
in the microfluidic chamber. At the resonance frequencies, the
behavior of the different tracer particle solutions was observed.
Between successive recordings the chip was flushed to assure
homogeneous seeding. Furthermore, to make sure that only particle
motion caused by acoustic forces were recorded, no external flow
was applied during measurements.

\subsection{Micro-PIV considerations}
\label{sec:MicroPIV}

In micro-PIV tracer particles are chosen for their ability to
truthfully follow the motion of the flow that is to be
investigated. Particles under the influence of an acoustic field
do no longer fulfil this criterium in all situations. Therefore,
extra caution and considerations have to be taken regarding what
movements are actually measured when applying micro-PIV for these
types of studies. Given that particle motion caused by thermal or
gravitational forces can be neglected, the main task is to
determine if particle motion is caused by acoustic radiation
forces, acoustic streaming or a combination of the two. In this
study, this problem was tackled by applying three tracer solutions
with different physical properties.

\begin{table}
\caption{\label{tab:chipgeom} The geometrical parameters of the
fabricated microfluidic silicon-pyrex chip.}
\begin{tabular}{lcr|@{\hspace*{2mm}}lcr} \hline
 chip length & $L^{{}}_0$ & 49~mm  &
 silicon thickness & $h^{{}}_s$ & 500~$\mu$m \\
 chip width & $w^{{}}_0$ & 15~mm  &
 pyrex thickness & $h^{{}}_p$ & 500~$\mu$m \\
 channel length & $L^{{}}_c$ & 26~mm  &
 chamber height & $h$ & 200~$\mu$m \\
 channel width & $w^{{}}_c$ & 400~$\mu$m &
 chamber width &  $w$ & 2~mm \\\hline
\end{tabular}
\end{table}

Typically, the large polyamide particles are more strongly affected
by the acoustic radiation forces than by the forces due to acoustic
streaming of the surrounding water. In contrast, since the acoustic
radiation force scales with the volume of the particle, the small
polystyrene particles will follow the motion of the water, if
relatively strong acoustic streaming is present. However, there is
no simple relation between the two forces, and for an arbitrary
frequency and geometry one can be strong whereas the other is not,
and vice versa. Therefore, in order to determine whether particle
motion is caused by acoustic radiation forces or acoustic streaming
it is necessary to utilize the dependance of the acoustic radiation
forces on the compressibility of the particle.

The polymer particles will move towards the pressure nodes since
their compressibility is smaller than that of water. The opposite is
true for the lipid particles in milk: their compressibility is
larger than that of water, and consequently they will move towards
pressure antinodes. Like the small polystyrene particles, the lipid
particles we used were small enough to typically follow the net
acoustic streaming flow of the water. Thus, if similar motion is
recorded with two types of tracers with different compressibilities
compared to the medium, the acoustic radiation forces can be ruled
out as cause of the motion. As an alternative or complementary
technique to micro-PIV measurements, fluorescein can be used to
investigate acoustic streaming. A summary of the acoustic behavior
of the different particles used in this study, and some other bodies
that are common in microfluidic applications, is given in
Table~\ref{tab:particles}.

The speed of sound $c$ in water has a significant dependence on
temperature $T$ given by the large derivative $\partial c /
\partial T \simeq 4$~m\,s$^{-1}$K$^{-1}$. All tracer fluids were
therefore kept at room temperature, so that the temperature was not
changed when the microchip was flushed during tracer particle
exchange. The microchips used in this study are comparable in size
and mode of actuation to those used for ultrasonic agitation in a
study by Bengtsson and Laurell~\cite{Bengtsson2004}. They performed
sensitive temperature measurements on the reactor outlet, where no
temperature increase caused by the acoustic power could be detected.
In our study, the piezo-actuator was run at a moderate power-level
and only for the short intervals during recordings (typically less
than one second). Therefore, it can be ruled out that heating from
the piezo-actuator would have any measurable impact on the
measurements.

\begin{table}
\caption{\label{tab:particles} The susceptibility to acoustic
radiation forces for the particles used in this study, as well as
for some other particles common to microfluidic applications.}
\begin{tabular}{l|@{\hspace*{2mm}}c|@{\hspace*{2mm}}c} \hline
 tracer type& force & direction \\ \hline
 beads (1 $µ$m) & weak & nodes \\
 beads (5 $µ$m) & strong & nodes \\
 red blood cells & strong & nodes \\
 milk particles & weak & anti-nodes \\
 large micelles & strong & anti-nodes \\
 fluorescein & none & - \\
\hline
\end{tabular}
\end{table}

One important factor, which needs to be accounted for when applying
micro-PIV on systems affected by acoustic forces, is that the local
seeding density will be distorted during actuation. This is normally
not a problem when measuring on particle motion caused solely by
acoustic streaming, as this motion generally will be of a
circulating nature. On the other hand, in the case of particle
motion induced by acoustic radiation forces, it will typically lead
to total expulsion of particles from certain regions into others. If
PIV-vector statistics is applied, only the first few image-pairs
recorded after piezo-actuation has been initiated can be used, and
in this study, images from a number of consecutively recorded sets
have been used for averaging. Moreover, in the case of scanning, or
mapping, techniques the expulsion of particles is especially
problematic, as the seeding conditions in the device, or chamber,
need to be restored for each measurement position. Also, the
conditions may change during these lengthy recordings, leading to
results that are difficult to interpret.

\begin{table}
\caption{\label{tab:acoustparam} The acoustic material parameters of
the microsystem at 20~$^\circ$C: sound velocities $c^{{}}_i$ and
densities $\rho^{{}}_i$ from the CRC Handbook of Chemistry and
Physics.}
\begin{tabular}{l|@{\hspace*{2mm}}r|@{\hspace*{2mm}}rr@{$\;$}} \hline
 material & speed of sound & \multicolumn{2}{c}{density} \\ \hline
 water   & $c^{{}}_w = $ 1483~m/s & $\rho^{{}}_w = $ & 998~kg/m$^3$ \\
 silicon & $c^{{}}_s = $ 8490~m/s & $\rho^{{}}_s = $ & 2331~kg/m$^3$ \\
 pyrex   & $c^{{}}_p = $ 5640~m/s & $\rho^{{}}_p = $ & 2230~kg/m$^3$ \\ \hline
\end{tabular}
\end{table}

The acoustic resonances in low attenuation piezo-actuated
microfluidic devices are formed over the whole devices, and they are
also depending on the geometry of the whole device. As a
consequence, there will typically be amplitude fluctuations over the
devices, due to unwanted artifacts, or deliberate designs.
Therefore, when investigating acoustic resonances, and the influence
caused by different modifications to the sample, it is important to
study the effects globally. If the acoustic effects are only
measured in a part of the device, this kind of information will not
be yielded, independently on how detailed the flow is mapped within
that region. Therefore, we suggest full-image micro-PIV for the
investigation of acoustic resonances in microfluidic devices.

In this study, emphasis has been put on how to present the measured
data in such a way that still images and PIV-vector plots give the
best illustration of the transient particle motion caused by the
acoustic forces. To achieve this, we have chosen to superimpose the
PIV-vector plots of the initial transient velocities on top of the
pictures of the steady-state patterns of the particles obtained
after a few seconds of actuation. After longer actuation times,
secondary patterns will form, so images taken at this point can give
a false impression of the particle motion. This method of combining
the transient PIV-vector plots and steady-state pictures has shown
useful when comparing numerical simulations with micro-PIV
measurements, especially for measuring amplitude fluctuations across
the structures, and when discriminating between different numerical
models. This will be demonstrated in Sec.~\ref{sec:Results}.

\section{Numerical simulations}
\label{sec:NumSim}

In the experiments, the acoustic pressure field, which is
superimposed on the ambient constant pressure, is driven by a
harmonically oscillating piezo-actuator, i.e., the time-dependence
can be described as $\cos(\omega t)$. In this work, we focus on
the acoustic resonances where the response of the bead solution is
particularly strong. As the attenuation of the acoustic waves is
relatively small, we can approximate the actual
frequency-broadened acoustic resonances of the driven system by
the infinitely sharp eigenmodes of the isolated dissipationless
chip.

\begin{figure}
\includegraphics[width=8.0 cm]{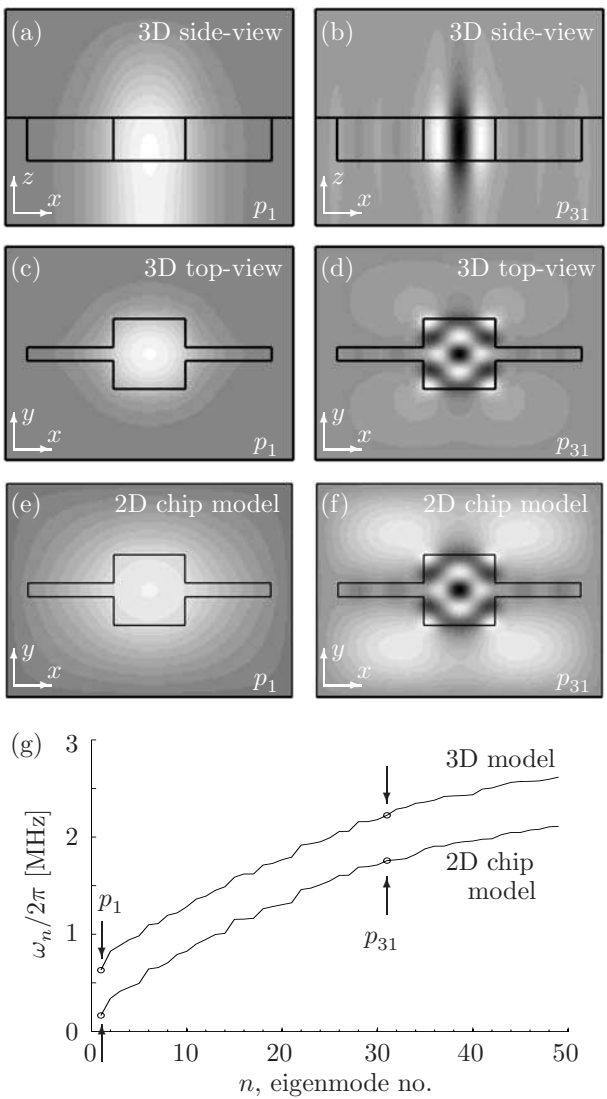}
\caption{Numerical simulations of the pressure eigenmodes
$p^{{}}_n(x,y,z)$ shown in gray-scale plots. (a) and (b) 3D model:
side-view ($xz$-plane) of $p^{{}}_1$ and $p^{{}}_{31}$,
respectively. (c) and (d) 3D model: top-view ($xy$-plane) of
$p^{{}}_1$ and $p^{{}}_{31}$, respectively. (e) and (f) 2D chip
model: top-view ($xy$-plane) of $p^{{}}_1$ and $p^{{}}_{31}$,
respectively. (g) The eigenfrequencies $\omega^{{}}_n/2\pi$ versus
mode number $n$ for the 3D model and the 2D chip model.}
\label{fig:Simul3D2D}
\end{figure}

The pressure eigenmodes $\pn(x,y,z)\cos(\on t)$, labelled  by an
integer index $n$, and the angular eigenfrequencies or resonance
frequencies $\on$ are found as solutions to the Helmholtz eigenvalue
equation $\nabla^2 \pn = -(\onsqr/c^{2_{}}_i)\,\pn$, where the index
$i$ is referring to the three material domains of silicon, water and
glass in the chip. The boundary conditions at the outer edges of the
system are given by the soft-wall condition $\pn = 0$ except at the
bottom plane, where a hard-wall condition $\nnn\cdot\nablabf \pn =
0$ is chosen to mimic the piezo-actuator which fixes the velocity of
the wall. At the internal interfaces between the different material
regions the boundary conditions are continuity of the pressure $\pn$
as well as of the wall-velocity. The latter is ensured by continuity
of the field $(1/\rho^{{}}_i)\nablabf \pn$. A list of the acoustic
material parameters, i.e., sound velocities $c^{{}}_i$ and densities
$\rho^{{}}_i$, is given in Table~\ref{tab:acoustparam}.

The Helmholtz equation was solved numerically using the COMSOL
finite element method software. However, the large aspect ratio of
the flat device made it impossible to simulate the actual device in
3D due to limited computer memory. We therefore investigated the
possibility of making 2D simulations. The rationale for doing this
is that the total height of the chip is only 1~mm. Given a weighted
average speed of sound in the silicon-glass chip of $6900$~m/s, the
wavelength of a wave at the highest used frequency $f = 2.5$~MHz is
3~mm and thus three times the chip height. Similarly, at the same
frequency the wavelength in water is 0.6~mm or three times the
chamber height. Consequently, there is not room enough for even half
a standing wave in the vertical direction neither in the water
filled chamber nor in the silicon-glass chip.

The first step towards a more rigorous justification for doing 2D
simulations was to make a smaller 3D version of the system geometry.
While keeping all the correct height measures as well as the chamber
width as listed in Table~\ref{tab:chipgeom}, we shrunk the planar
dimensions of the surrounding chip to $L^{{}}_0 = 8$~mm, $w^{{}}_0 =
6$~mm and $L^{{}}_c = 6.8$~mm. With this reduced geometry we could
carry out the full 3D simulations, and the results thereof confirmed
that the variations in the vertical $z$-direction of the 3D
eigenmodes were modest, see the $xz$-plane plots of
Figs.~\ref{fig:Simul3D2D}(a) and~(b). A 2D simulation was then
carried out for the horizontal $xy$ center-plane of the chamber,
i.e., a 2D water-filled area surrounded by a 2D silicon region.
Comparing the 50 lowest 3D and 2D eigenmodes gave the following
results: (1) in the horizontal $xy$ center-plane of the chamber the
3D eigenmodes agreed with the 2D eigenmodes, see
Figs.~\ref{fig:Simul3D2D}(c--f); (2) due to the lack of the
$z$-dependence in the Laplacian of the 2D Helmholtz equation, the 2D
eigenfrequencies were systematically smaller than the 3D
eigenfrequencies, see Fig.~\ref{fig:Simul3D2D}(g). It has thus been
justified to simulate the experimentally observed resonances by 2D
eigenmodes in the horizontal $xy$ center-plane of the chamber. This
we denote the \emph{2D chip model}.

\begin{figure}
\includegraphics[width=8.0 cm]{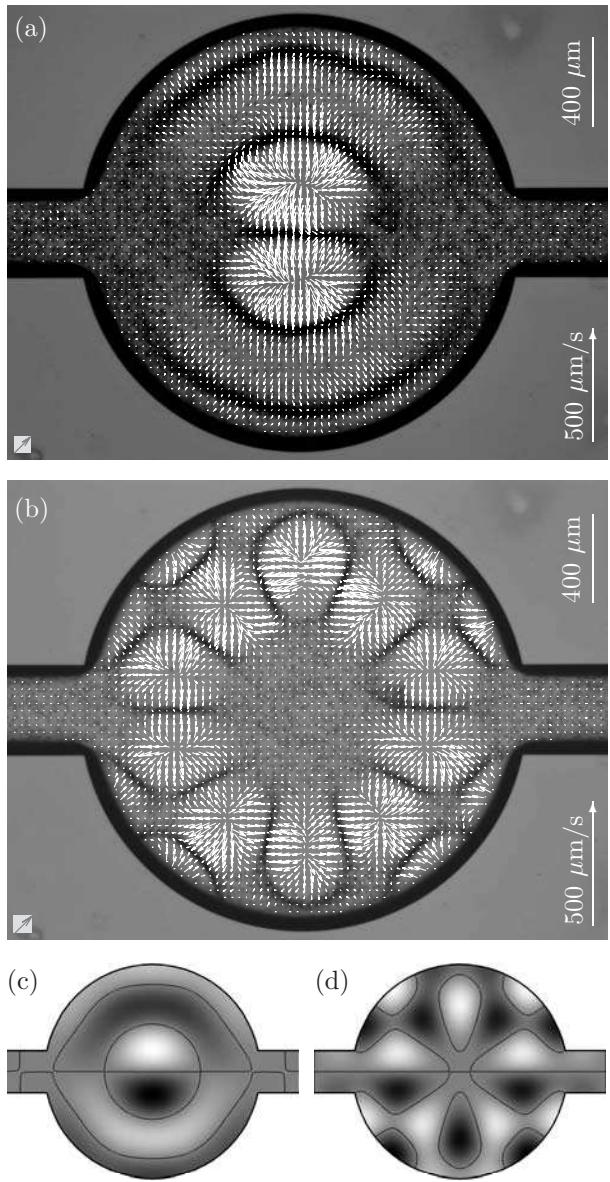}
\caption{Acoustic radiation force. (a) Experiments on 5~$\mu$m beads
at the 1.936~MHz acoustic resonance. The white PIV-vectors indicate
the initial bead velocities pointing away from pressure anti-nodes
immediately after the piezo-actuation is applied. The picture
underneath the PIV-vector plot shows the particles (black) gathered
at the pressure nodal lines 3 seconds later. (b) As in panel (a) but
now at 2.417~MHz. (c) and (d) Gray-scale plots of numerical
simulations in the 2D chamber model of the corresponding acoustic
pressure eigenmodes. Nodal lines are shown in black.}
\label{fig:AcoustRadForce}
\end{figure}

Due to the small acoustic impedance ratio $(\rho^{{}}_w
c^{{}}_w)/(\rho^{{}}_s c^{{}}_s) = 0.08$ between silicon and water,
the simulations could be simplified even further. As demonstrated in
Figs.~\ref{fig:AcoustRadForce}(c) and~(d), it suffices to find the
eigenmodes of the chamber itself using hard-wall boundary conditions
along its edges, except at the very ends of the inlet channels where
soft-wall boundary conditions are employed to mimic in- and outlets.
This we will refer to as the \emph{2D chamber model}.

\section{Results and discussion}
\label{sec:Results}

We have measured the flow response to the acoustic actuation in the
frequency range from 0.5 to 2.5~MHz paying special attention to the
strong responses corresponding to acoustic resonances. More than 30
of such resonances have been detected, but we present only a few,
which we find to be representative for the method and the problems
associated with acoustics in microfluidics.

The most important results are the full-image micro-PIV analyses.
For these, two types of experimental results are presented. One type
are the PIV-vector plots (white arrows) of the motion of the tracer
particles, in most cases corresponding to the transient motion
immediately after the onset of the acoustic piezo-actuation. The
other type are micrographs of the microfluidic chamber with the
steady-state particle patterns (often visible as narrow black bands)
obtained after a few seconds of actuation. These two types of images
are superimposed to illustrate the relation between the initial
motion of the tracer beads and their final steady-state positions.

The full-image micro-PIV analysis illustrations are also accompanied
by the results of our numerical simulations in the form of
gray-scale plots of the pressure eigenmodes $p^{{}}_n(x,y,z)$. The
pressure antinodes appear as white (positive amplitude) and black
(negative amplitude) regions. The pressure nodal lines are shown as
thin black lines in the gray (small amplitude) regions.

Additionally, we show a close up measurement of a streaming vortex,
and provide a more in-depth comparison between the measured
velocities and the calculated body force.

\subsection{Acoustic radiation force}
\label{sec:AcoustRadForce} We first show results for the acoustic
resonances at 1.936 and 2.417~MHz in the circular chamber containing
large 5~$\mu$m tracer particles.

In Figs.~\ref{fig:AcoustRadForce}(a) and~(b) are shown the
measured transient PIV-vector plots superimposed on the
micrographs of the chamber with the static steady-state particle
patterns. The fact that the particles accumulate in static
patterns indicates that the dominant force on the tracer particles
is the acoustic radiation force, an observation also expected from
the relatively large size of the tracer particles. The matching
numerically calculated acoustic eigenmodes of the 2D chamber model
are shown in panels~(c) and~(d). It is noteworthy that even for
the complicated resonance pattern of panels~(b) and~(d), the
observed transient particle motion towards the steady-state
positions, and the static steady-state patterns themselves, are in
good agreement with the numerically calculated pressure nodal
lines. This demonstrates that even the simple 2D chamber model can
predict what kind of fluidic behavior will be observed in the
device. It also demonstrates that full-image micro-PIV analysis in
combination with images of transient particle motion is effective
in visualizing in-plane acoustic phenomena in micrometer-scale
devices.

\begin{figure}
\includegraphics[width=8.0 cm]{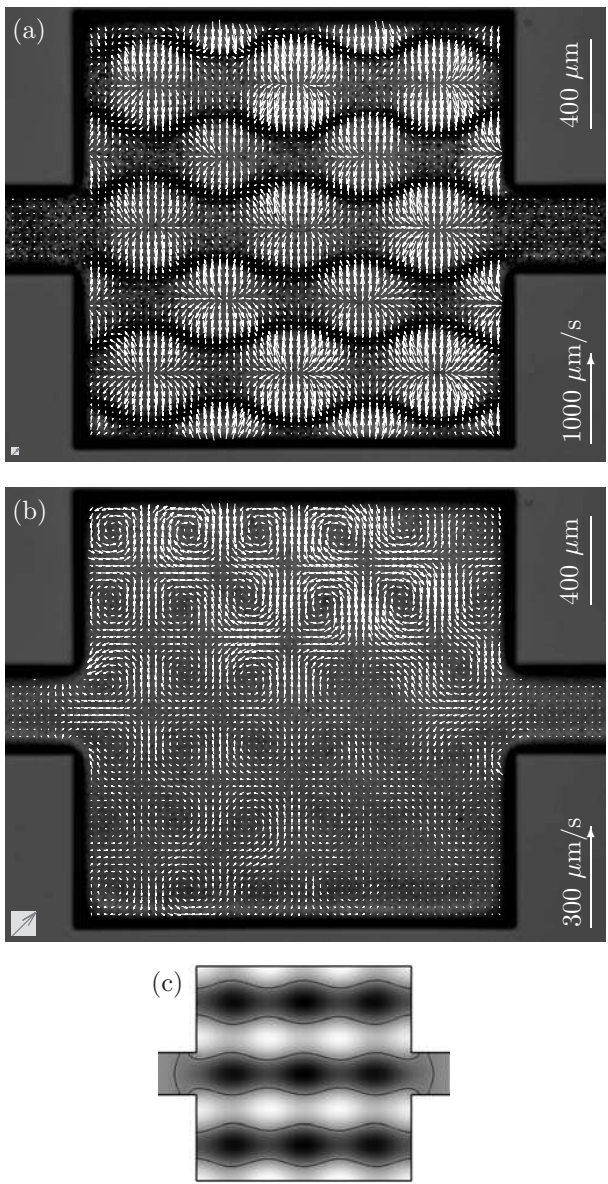}
\caption{Acoustic streaming and radiation forces at the 2.17~MHz
acoustic resonance. (a) Experiments on 5~$\mu$m beads similar to
Fig.~\ref{fig:AcoustRadForce}(a) showing that the acoustic radiation
force dominates for large particles. (b) Experiments on 1~$\mu$m
beads. Acoustic streaming dominates and the small beads act as
tracers for the motion of the liquid. The resulting vortex structure
in the flow-field prevents particle accumulation at the pressure
nodes. (c) Gray-scale plot of numerical simulation in the 2D chamber
model of the corresponding acoustic pressure eigenmode. Nodal lines
are shown in black.} \label{fig:AcoustStreaming}
\end{figure}

\subsection{Acoustic streaming}
\label{sec:AcoustStreaming} To illustrate the difference between the
acoustic radiation force and acoustic streaming, we now turn to the
acoustic resonance at 2.17~MHz in the square chamber containing
large 5~$\mu$m beads and small 1~$\mu$m beads as shown in
Figs.~\ref{fig:AcoustStreaming}(a) and~(b), respectively.

When micro-PIV is applied to investigate acoustic effects in
microfluidic chambers, the simultaneous presence of both acoustic
radiation forces and acoustic streaming needs to be taken into
account. For the large beads in Fig.~\ref{fig:AcoustStreaming}(a)
the acoustic radiation force dominates exactly as in
Figs.~\ref{fig:AcoustRadForce}(a) and~(b), which results in particle
accumulation at the pressure nodal lines. However, as shown in
Fig.~\ref{fig:AcoustStreaming}(b) reduction of the particle volume
by a factor of 125 leads to a qualitative change in the response.
The motion of the smaller particles is dominated by the acoustic
streaming of the water, and it manifests itself as a 6$\times$6
pattern of vortices. The same 6$\times$6 pattern was found by
full-image micro-PIV when diluted milk was used as tracer solution,
and also by optical inspection with a fluorescein solution in the
chamber (data not shown). All three experimental results strongly
support the interpretation that the 6$\times$6 vortex pattern is
caused by acoustic streaming.

In Fig.~\ref{fig:AcoustStreaming}(b) is also seen a pronounced
inhomogeneity in the strength of the vortices across the
microfluidic chamber. This effect cannot be ascribed to the geometry
of the chamber, but is probably due to either a geometric top-bottom
asymmetry in the entire chip (similar to the left-right asymmetry
discussed in Sec.~\ref{sec:GeomAsym}), or to an inhomogeneous
coupling between the piezo-actuator and the silicon chip. If the
frequency is shifted slightly in the vicinity of 2.17 MHz, the same
vortex pattern will still be visible, but the strength distribution
between the vortices will be altered. When investigating acoustic
phenomena the advantage of full-image micro-PIV compared to
partial-image micro-PIV is thus evident: partial-image micro-PIV
employed locally in a part of the chamber would not have shown the
symmetrical 6$\times$6 vortex pattern, nor would it supply us with
information of the inhomogeneity in strength for the same. Moreover,
since the same inhomogeneity is not seen in the acoustic radiation
force vector plot, this example shows that there is no direct
relation between the strength of the acoustic streaming and the
acoustic radiation force.

Finally, we note that our measurements show that the acoustic
radiation force on the large particles leads to a much larger
particle velocity than the acoustic streaming velocities of the
smaller particles.

Turning to the numerical simulation in the 2D chamber model of the
corresponding pressure eigenmode, shown in
Fig.~\ref{fig:AcoustStreaming}(c), we find good agreement with the
experimental results. The calculated pressure nodal lines correspond
well to the static steady-state particle patterns obtained with the
large tracer particles dominated by the acoustic radiation force.
Moreover, the calculated 3$\times$3 antinode pattern is also
consistent with the observed period-doubled 6$\times$6 vortex
pattern of the small tracer particles dominated by acoustic
streaming. The spatial period-doubling arises from the non-zero
time-average of the non-linear term in the Navier--Stokes equation
governing the attenuated acoustic flows leading to acoustic
streaming~\cite{Lighthill}.

\subsection{Effects of geometric asymmetries}
\label{sec:GeomAsym}

For the results presented so far the simple 2D chamber model
proved sufficient to interpret the experimental observations.
However, as explained already in Sec.~\ref{sec:NumSim} the
pressure eigenmodes are not confined to the chamber region but
fill the entire chip. The acoustic resonances even propagate in
all media (air and piezo-actuator) in contact with the chip. In
the following we show one example of asymmetric resonance patterns
that can only be explained by employing the more complete 2D chip
model or by introducing asymmetries in the 2D chamber model.

\begin{figure}
\includegraphics[width=8.0 cm]{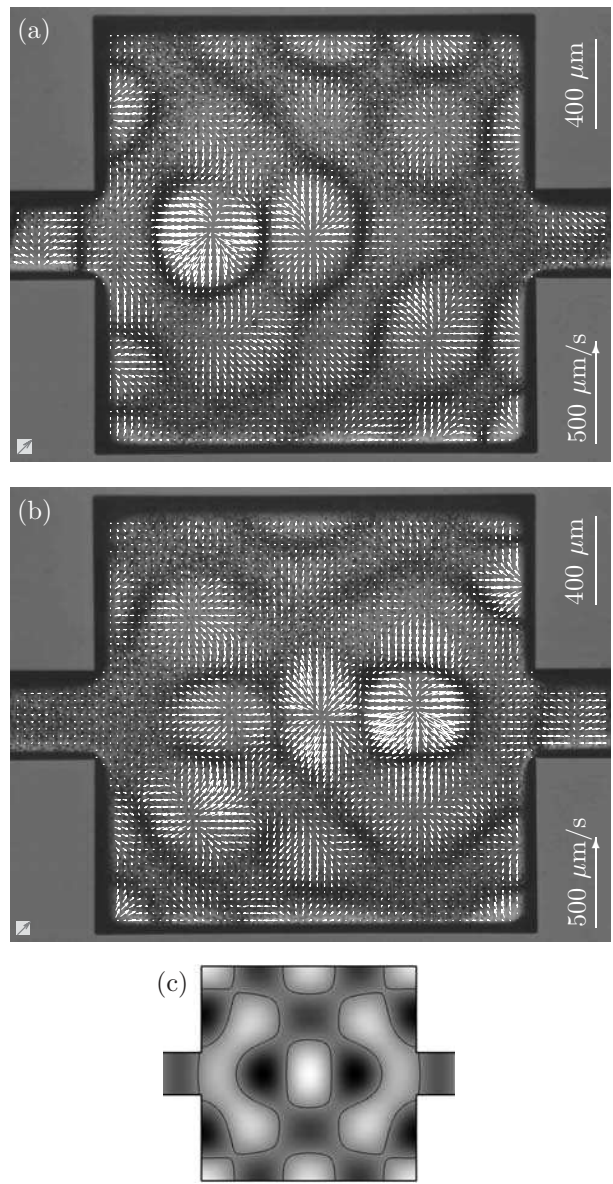}
\caption{Splitting of a two-fold degenerate acoustic resonance due
to geometrical asymmetry. (a) Acoustic radiation force as in
Fig.~\ref{fig:AcoustStreaming}(a) on 5~$\mu$m beads at the 2.06~MHz
resonance. (b) The closely related 2.08~MHz resonance for the same
system. (c) Gray-scale plot of numerical simulation in the
left-right symmetric 2D chamber model of the corresponding two-fold
degenerate, un-split, acoustic pressure eigenmode. Nodal lines are
shown in black.} \label{fig:AcoustRadAssym}
\end{figure}

\begin{figure}
\includegraphics[width=8.0 cm]{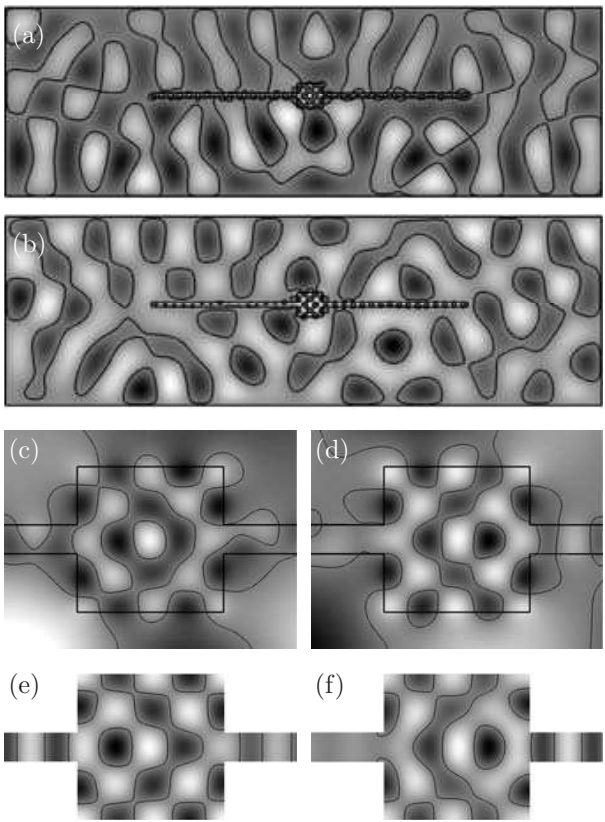}
\caption{(a) ad (b) Gray-scale plots of numerical simulations in
the 2D chip model of two closely spaced acoustic pressure
eigenmodes. The chamber is displaced 1~mm to the left of the
symmetry center of the chip thereby breaking the left-right
symmetry and splitting the two-fold eigenmode degeneracy. The
difference in eigenfrequency is only 1~kHz. (c) and (d) Closeups
of the chamber region showing the asymmetric eigenmodes similar to
the experimentally observed resonances seen in
Fig.~\ref{fig:AcoustRadAssym}. (e) and (f) Gray-scale plots of
numerically simulated pressure eigenmodes in the asymmetric 2D
chamber model, where the left lead is 1~mm shorter than the right
lead. The difference in eigenfrequency is 28~kHz, which is close
to the observed difference of 20~kHz in
Figs.~\ref{fig:AcoustRadAssym}(a) and~(b).}
\label{fig:2dChipModel}
\end{figure}

In Figs.~\ref{fig:AcoustRadAssym}(a) and~(b) we consider the square
chamber containing the large 5~$\mu$m beads at two nearby resonance
frequencies, 2.06 and~2.08~MHz. As before, the acoustic radiation
force dominates and the beads accumulate at the pressure nodal
lines. Note that the two patterns are similar, but that the first
has a higher amplitude on the left side, while the second has a
higher amplitude on the right side. Both resonance patterns are
similar to the acoustic pressure eigenmode shown in
Fig.~\ref{fig:AcoustRadAssym}(c), which is found by numerical
simulation using the 2D chamber model. However, since the chamber
itself is left-right symmetric, the calculated eigenmode is also
left-right symmetric, so to explain the observed asymmetry we have
to break the left-right symmetry in the theoretical model. We
investigate two ways of doing this: first, in the 2D chip model by
placing a symmetric chamber asymmetrically on the chip, and second,
in the 2D chamber model by letting the inlet channel have a
different length than the outlet channel.

In Figs.~\ref{fig:2dChipModel}(a-d) is shown the result of a
numerical simulation in the 2D chip model where the left-right
symmetry has been broken by displacing the chamber 1~mm left of the
symmetry center of the chip. This displacement corresponds to the
geometry of the actual chip used in the experiment. Panels~(a)
and~(b) show the entire chip while panels~(c) and~(d) are the
corresponding closeups of the chamber region. With this left-right
asymmetric geometry, we do find asymmetric solutions at nearby
frequencies that resemble the measured patterns:
Figs.~\ref{fig:2dChipModel}(c) and (d) correspond to
Figs.~\ref{fig:AcoustRadAssym}(a) and (b), respectively. In the
left-right symmetric case the left-right acoustic resonance is
two-fold degenerate, i.e., two different resonances have the same
frequency. When the symmetry is broken the two resonances are
affected differently: one gets a slightly higher eigenfrequency and
the other a slightly lower, i.e., a splitting of the two-fold
degenerate eigenfrequency into two non-degenerate nearly identical
eigenfrequencies. The two closely spaced eigenmodes of the
asymmetric 2D chamber model shown in Figs.~\ref{fig:2dChipModel}(e)
and~(f) also resemble the measured patterns in
Figs.~\ref{fig:AcoustRadAssym}(a) and (b). The calculated frequency
splitting is 28~kHz, which is in fair agreement with the measured
20~kHz.

Unquestionably, advanced models, like the chip model, are necessary
for more complete theoretical investigations of how different
factors contribute to the breaking of the symmetry of the simple
chamber model. Experimentally this effect could be studied by
measuring on a range of devices, with strictly controlled geometries
of both structures and substrates. We have only investigated two
devices, and special concern was not taken as to the uniformity of
the substrate. It is therefore not possible in the present study to
determine whether the observed symmetry breaking was due to
geometric asymmetries in the chip, in the chip-actuator coupling, or
in other parts of the system (such as air-bubbles trapped at the
fluidic inlet and outlet).

\subsection{Validation of method}

\begin{figure}
\includegraphics[width=8.0 cm]{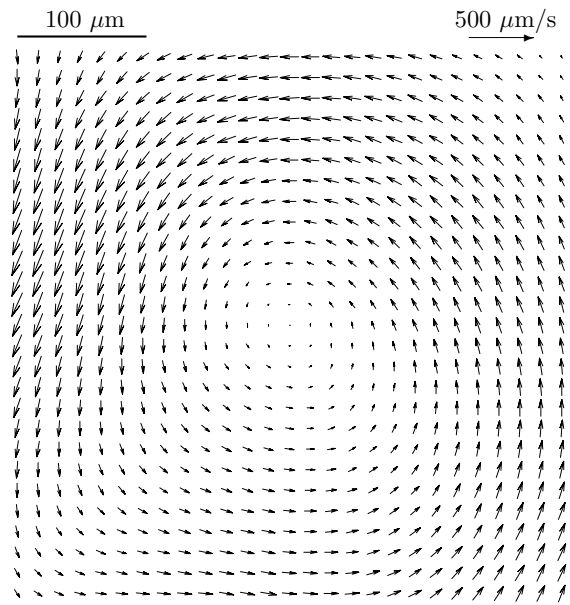}
\caption{Micro-PIV velocity vector plot of streaming motion in the
center of the square chamber at 2.17 MHz. Images were recorded with
a 20x objective and a 0.63x Tv-adapter, and milk was used as tracer
particles.} \label{fig:20xPIV}
\end{figure}

Fig.~\ref{fig:20xPIV} shows a micro-PIV vector plot of streaming
motion in the center of the square chamber at 2.17~MHz, recorded
with a 20x microscope objective. With this kind of recording,
detailed information of a section of the device can be obtained, but
it will not supply any information about the amplitude fluctuations
over the device, nor does it reveal the 6x6 vortex pattern as seen
in Fig.~\ref{fig:AcoustStreaming}(b). Clearly, more detailed
measurements of specific features are valuable, but for studies of
resonances in low attenuation microfluidic devices, full-image
recordings are of most importance.

\begin{figure}
\includegraphics[width=8.0 cm]{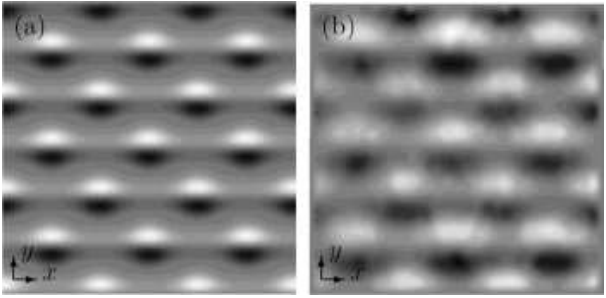}
\caption{(a) The force in $y$-direction calculated with COMSOL
finite element method software. (b) Scalar map of the velocity in
$y$-direction, measured with micro-PIV. } \label{fig:doubleplots}
\end{figure}

\begin{figure}
\includegraphics[width=8.0 cm]{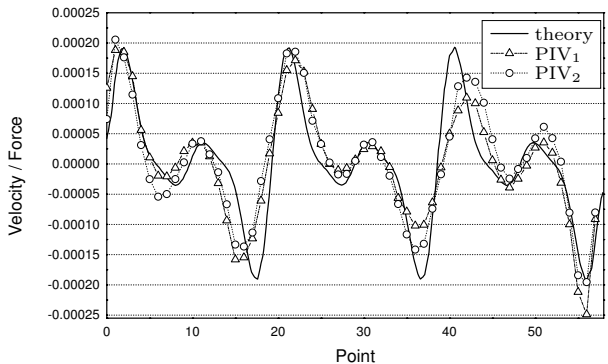}
\caption{Vertical cross-sectional plots of the velocity in the
$y$-direction (PIV$_1$) 330~$µ$m left and (PIV$_2$) 330~$µ$m right
of the center of the chamber. (theory) is the force in
$y$-direction, calculated in COMSOL with the amplitude as the only
fitting parameter.} \label{fig:comparegraph}
\end{figure}

By regarding the suspended particles as springs governed by Hooke's
law, we can estimate the acoustic radiation force $\mathbf{F}_{ac}$,
from the potential elastic energy. We find $\mathbf{F}=-\tau
p^{{}}_2 \boldsymbol{\nabla} p^{{}}_2$, where $\tau$ is a constant
parameter for each kind of particles, and $p^{{}}_2$ is the
time-averaged second-order pressure field. As $\tau$ is an unknown
positive constant for blood-cell like particles, the amplitude
becomes a fitting parameter. Assuming the particles move in a
quasi-stationary steady state, we can directly compare calculated
force patterns to measured velocity patterns. Such a comparison is
seen in Fig.~\ref{fig:doubleplots}, where the calculated force
pattern is compared with a scalar map of the velocity in
$y$-direction, extruded from the measurement presented in
Fig.~\ref{fig:AcoustStreaming}(a). A comparison between the two is
also seen in Fig.~\ref{fig:comparegraph}, where two vertical
cross-sectional views, each located 330~$µ$m away from the center of
the chamber, are compared with the theoretical estimate. Both
micro-PIV velocity plots show a good agreement with the calculated
forces, and the fluctuations in amplitude over the device can be
seen by comparing the two micro-PIV velocity plots with each other.

%
%
%

\section{Conclusion}
\label{sec:Conclusion}

Using full-image micro-PIV we have made direct observations of the
acoustic resonances in piezo-actuated, flat microfluidic chambers
containing various tracer particles.

Depending on the size of the tracer particles either the acoustic
radiation force or acoustic streaming of the solvent dominates their
motion. Large particles are dominated by the acoustic radiation
force that pushes them to the static pressure nodal lines, while
small particles are dominated by acoustic streaming and end up
forming steady-state vortex patterns. However, for an arbitrary
frequency and geometry one of the forces can be strong whereas the
other is not, and it is therefore always necessary to apply more
than one tracer solution in order to determine which forces are
present.

The observed acoustic resonances correspond to the pressure
eigenmodes found by numerical simulation of 2D models of the system.
The symmetric patterns can be explained by using the simple 2D
chamber model, while asymmetric patterns can be explained by using
the more complete 2D chip model taking into account the geometric
asymmetries of the surrounding chip, or in special cases, by an
asymmetric 2D chamber model.

We have demonstrated that full-image micro-PIV is a useful tool for
complete characterization of the in-plane acoustically induced
motion in piezo-actuated microfluidic chambers.
\\
\section*{Acknowledgement}
SMS was supported through Copenhagen Graduate School of Nanoscience
and Nanotechnology, in a collaboration between Dantec Dynamics A/S,
and MIC, Technical University of Denmark.


\begin{thebibliography}{99}

\bibitem{King1934}
L.V. King,
\emph{Proc. R. Soc. London, Ser. A}, 1934, \textbf{147}, 212.

\bibitem{Yosioka1955}
K.~Yosioka and Y.~Kawasima,
\emph{Acustica}, 1955, \textbf{5}, 167.

\bibitem{Gorkov1962}
L.P. Gorkov,
\emph{Sov. Phys. Doklady}, \textbf{6}, 1962, 773.

\bibitem{Rayleigh1883}
Lord Rayleigh,
\emph{Proc. R. Soc. London}, 1883, \textbf{36}, 10.

\bibitem{Riley2001}
N. Riley,
\emph{Annu. Rev. Fluid Mech.}, 2001, \textbf{33}, 43.

\bibitem{Yasuda1995}
K. Yasuda, S. Umemura, and K. Takeda,
\emph{Jpn. J. Appl. Phys., Part 1}, 1995, \textbf{34}, 2715.

\bibitem{Yasuda1996}
K. Yasuda, K. Takeda, and S. Umemura,
\emph{Jpn. J. Appl. Phys., Part 1}, 1996, \textbf{35}, 3295.

\bibitem{Zhu1998}
X. Zhu and E. S. Kim,
\emph{Sens. Actuators, A}, 1998, \textbf{66}, 355.

\bibitem{Yang2001}
Z. Yang, S. Matsumoto, H. Goto, M. Matsumoto, and R. Maeda,
\emph{Sens. Actuators A}, 2001, \textbf{93}, 266.

\bibitem{Liu2003a}
R. H. Liu, R. Lenigk, R. L. Druyor-Sanchez, J. Yang, and P.
  Grodzinski,
\emph{Anal. Chem.}, 2003, \textbf{75}, 1911.

\bibitem{Rife2000}
J. C. Rife, M. I. Bell, J. S. Horwitz, M. N. Kabler, R. C. Y.
Auyeung, and W. J. Kim,
\emph{Sens. Actuators A}, 2000, \textbf{86}, 135.

\bibitem{Andersson2001}
H. Andersson, W. van der Wijngaart, P. Nilsson, P. Enoksson, and
  G. Stemme,
\emph{Sens. Actuators B}, 2001 \textbf{72}, 259.

\bibitem{Saito2002}
M. Saito, N. Kitamura, and M. Terauchi,
\emph{J. Appl. Phys.}, 2002, \textbf{92}, 7581.

\bibitem{Lilliehorn2005}
T. Lilliehorn, U. Simu, M. Nilsson, M. Almqvist, T. Stepinski, T.
Laurell, J. Nilsson, and S. Johansson,
\emph{Ultrasonics}, 2005, \textbf{43}, 293.

\bibitem{Nilsson2004}
A. Nilsson, F. Petersson, H. Jonsson, and T. Laurell,
\emph{Lab Chip}, 2004, \textbf{4}, 131.

\bibitem{Jonsson2004}
H. Jonsson, C. Holm, A. Nilsson, F. Petersson, P. Johnsson, and T.
Laurell,
\emph{Ann. Thoracic Surgery}, 2004, \textbf{78}, 1572.

\bibitem{Li2004}
H. Li and T. Kenny,
\emph{Conf. Proc. 26 Ann. Int. Conf. IEEE Engineering in Medicne
and Biology}, 2004 \textbf{3}, 2631 Vol.4.

\bibitem{Wiklund2003}
M. Wiklund, P. Spegel, S. Nilsson and H. M. Hertz,
\emph{Ultrasonics}, 2003, \textbf{41}, 329.

\bibitem{Lutz2003}
B. R. Lutz, J. Chen, and D. T. Schwartz,
\emph{PNAS}, 2003, \textbf{100}, 4395.

\bibitem{Lutz2005}
B. R. Lutz, J. Chen, and D. T. Schwartz,
\emph{Phys. Fluids}, 2005, \textbf{17}, 1.

\bibitem{Santiago1998}
J. G. Santiago, S. T. Wereley, C. D. Meinhart, D. J. Beebe, and R.
J. Adrian,
\emph{Exp. Fluids}, 1998, \textbf{25}, 316.

\bibitem{Spengler2000a}
J. Spengler and M. Jekel,
\emph{Ultrasonics}, 2000, \textbf{38}, 624.

\bibitem{Spengler2003}
J.F. Spengler, W.T. Coakley, and K.T. Christensen,
\emph{AIChE J.}, 2003, \textbf{49}, 2773.

\bibitem{Kuznetsova2004}
L. A. Kuznetsova and W. T. Coakley,
\emph{J. Acoust. Soc. Am.}, 2004, \textbf{116}, 1956.

\bibitem{Jang2005}
L.S. Jang, S.H. Chao, M.R. Holl, and D.R. Meldrum,
\emph{Sens. Actuators A}, 2005, \textbf{122}, 141.

\bibitem{Manasseh2006}
R. Manasseh, K. Petkovic-Duran, P. Tho, Y. Zhu, and A. Ooi,
\emph{BioMEMS and Nanotechnology II, Progress in Biomedical Optics
and Imaging - Proceedings of SPIE}, 2006, \textbf{6036}.

\bibitem{Sundin2007}
S. M. Sundin, C. H. Westergaard, H. Bruus, and J. P. Kutter,
\emph{Exp. Fluids}, 2007, submitted.

\bibitem{Bengtsson2004}
M. Bengtsson and T. Laurell,
\emph{Anal. Chem.}, 2004, \textbf{378}, 1716.

\bibitem{Lighthill}
J. Lighthill, \emph{Waves in fluid}, Cambridge University Press,
Cambridge, 2005.



\end{thebibliography}



\end{document}